\documentclass[showpacs,twocolumn,pra,aps,groupedaddress]{revtex4-1}

\usepackage{amsmath}
\usepackage{amssymb}
\usepackage{maybemath}
\usepackage{xcolor}
\usepackage{graphicx}
\usepackage{bm}
\usepackage{comment}
\usepackage{braket}
\usepackage{mathrsfs}
\usepackage{tikz}

\usetikzlibrary{decorations.pathmorphing}
\usetikzlibrary{shapes}

\usepackage{dcolumn}
\newcolumntype{.}{D{.}{.}{-1}}

\usepackage[%
colorlinks=true,
urlcolor=blue,
linkcolor=blue,
citecolor=blue
]{hyperref}

\makeatletter

\newcommand{\Rmnum}[1]{\expandafter\@slowromancap\romannumeral #1@}
\makeatother

\def\idmx{1\!{\rm l}}

\begin{document}

\title{Resonance fluorescence in the resolvent operator formalism}

\author{V. Debierre}
\affiliation{Max Planck Institute for Nuclear Physics, Saupfercheckweg 1, 69117 Heidelberg}
\author{Z. Harman}
\affiliation{Max Planck Institute for Nuclear Physics, Saupfercheckweg 1, 69117 Heidelberg}

\begin{abstract}
The Mollow spectrum for the light scattered by a driven two-level atom is derived in the resolvent operator formalism. The derivation is based on the construction of a master equation from the resolvent operator of the atom-field system. We show that the natural linewidth of the excited atomic level remains essentially unmodified, to a very good level of approximation, even in the strong-field regime, where Rabi flopping becomes relevant inside the self-energy loop that yields the linewidth. This ensures that the obtained master equation and the spectrum derived matches that of Mollow.
\end{abstract}


\maketitle

\newpage

\section{Introduction} \label{sec:Intro}

It is well known that the spectrum of the light scattered by a two-level system---usually an atom---exhibits very different profiles in different intensity regimes \cite{Mollow,FoxMulder,JeEvHaKeLetter,JeKeLong,EvJeKe}. This is best seen by looking at the incoherently scattered fraction of the incoming light, which is subdominant at low intensities and dominant at high intensities. Mollow has shown \cite{Mollow} that the incoherent part takes over when the modulus $\left|\Omega\right|$ of the Rabi frequency of the light-atom interaction becomes larger than the natural linewidth $\Gamma$ of the excited atomic state. For $\left|\Omega\right|\ll\Gamma$, the (subdominant) incoherent spectrum features a single Lorentzian peak around the resonance frequency $\omega_0$ of the atom (or, when some detuning $\Delta\equiv\omega_L-\omega_0\neq0$ is introduced between the atomic transition and the incoming light, two peaks at frequencies $\omega_0$ and $2\omega_L-\omega_0$, with $\omega_L$ the frequency of the laser). For $\left|\Omega\right|\gg\Gamma$, the (dominant) incoherent spectrum features the three notorious Mollow peaks, at frequencies $\omega_L$ and $\omega_L\pm\left|\Omega\right|$ (or, in the presence of detuning $\omega_L$ and $\omega_L\pm\sqrt{\left|\Omega\right|^2+\Delta^2}$). The usual derivation of the spectrum of the scattered light is based on a master equation that describes the evolution of the atomic density matrix under the influence of a classical driving field \cite{Mollow,FoxMulder}. A quantum electrodynamical (QED) derivation of the spectrum is possible in the low-intensity case, for example along the lines of the treatment given in Chapter~3 of \cite{CohenQED2}. In this case, it can be assumed that a single photon is absorbed and reemitted by the atom. However, in the high-intensity (strong field) case, this is no longer valid. Our aim is to extend the QED treatment to the strong field case.

Given the large variety of existing spectroscopic applications of QED, it is surprising that the description of the interaction of bound systems with light are, except for the simplest cases such as single-photon processes, almost exclusively formulated in the framework of master or Bloch equations \cite{KiffnerMacoveiLong,CohenQED2}, or by semi-classical methods such as the Weisskopf-Wigner model \cite{FoxMulder,CohenQED2}, and is not based on the foundations of QED. Multiphoton processes in particular are usually described by effective models such as master equations, for instance, in the so-called dressed atom approach, or by optical Bloch equations. Quantum-electrodynamic perturbation theory has been first applied by Low \cite{LowLineShape} to explain the natural line shape of an atom resonantly excited by a weak light field. He showed that the Lorentzian spectral line can be originated from an infinite series expansion of the photon scattering matrix in terms of Feynman diagrams with self-energy loop corrections to the intermediate state of the process. The case of overlapping resonances was later considered in a Green's function approach \cite{ShabaevFluo} and in an $S$-matrix approach \cite{RelatEffects}. In our efforts to rederive the Mollow spectrum, we lay the foundations of a framework that allows one to go beyond the approximation of a weak external field, and extend perturbation theory to the case when several real photons may be absorbed and reemitted during the process. These elementary multiphoton processes provide the QED interpretation of the Rabi oscillation phenomenon, which is a cornerstone of many laser spectroscopic and quantum optical applications \cite{CohenQED2,FoxMulder}.

A formalism based on QED is anticipated to be particularly important in the case of heavy systems. The laser spectroscopic study of atoms and ions with higher nuclear charges has recently been enabled by the construction of x-ray free electron laser facilities (see \emph{e.g.} \cite{SACLA}), allowing one to address important questions of astrophysics, such as \emph{e.g.} the strength of certain resonance lines in highly charged iron ions used for temperature determination of distant stars from spectra recorded by x-ray space observatories \cite{Bernitt}. It has been put forward \cite{OreshkinaAstro} that strong-field processes, such as can be investigated with the formalism developed in the current work, provide a possible resolution of discrepancies in laboratory astrophysics measurements with x-ray free-electron lasers \cite{Bernitt}.

We recall the definition of the spectral density of the scattered field in Sec.~\ref{sec:ScatField}, and Mollow's master-equation analysis of the problem in Sec.~\ref{sec:Mollow}. We derive rigorously the master equation from a resolvent operator formalism in Sec.~\ref{sec:ResoForma}, and conclude in Sec.~\ref{sec:Ccl}.

\section{The scattered field} \label{sec:ScatField}

The spectrum of the electric field scattered by a two-level atom is given by \cite{Titulaer,FoxMulder}
\begin{equation} \label{eq:ScatSpec}
\begin{aligned} [b]
G_{jk}\left(\omega,\mathbf{x}_{\mathrm{d}}\right)&\propto\int_0^{+\infty}\mathrm{d}t\,\mathrm{e}^{\mathrm{i}\omega t}\left\langle\hat{E}_j^-\left(\mathbf{x}_{\mathrm{d}},0\right)\hat{E}_k^+\left(\mathbf{x}_{\mathrm{d}},0\right)\right\rangle\\
&\propto F_{jk}\left(\mathbf{x}_{\mathrm{d}}\right)\int_0^{+\infty}\mathrm{d}t\,\mathrm{e}^{\mathrm{i}\omega t}\left\langle\hat{\sigma}_+\left(0\right)\hat{\sigma}_-\left(t\right)\right\rangle\\
&\propto F_{jk}\left(\mathbf{x}_{\mathrm{d}}\right)S\left(\omega\right),
\end{aligned}
\end{equation}
where $\hat{\sigma}_\pm\left(t\right)$ is the raising/lowering operator for the two-level atom at time $t$, in the Heisenberg picture, and $\mathbf{x}_{\mathrm{d}}$ is the position of the light detector on which the scattered field impinges. It is assumed that the detector sits in the far-field region of emission, where the emitted electric field varies as $\left\langle\mathbf{x}_{\mathrm{d}}\right\rangle^{-1}$. The prefactor $F_{jk}$, as such, is a trivial spatial factor that varies as $\left\langle\mathbf{x}_{\mathrm{d}}\right\rangle^{-2}$. The focus is to be put on the time-integral $S\left(\omega\right)$, which is the difficult part to compute. The angle brackets in (\ref{eq:ScatSpec}) refer to the expectation value taken over the state of the atom, which is described by a density matrix. As such, we see that it is the dynamics of the two-level atom, under the influence of the driving field, that determines the spectrum of the scattered electromagnetic field.

\section{Quantum optical treatment: The Mollow spectrum} \label{sec:Mollow}

Mollow's solution \cite{Mollow} to the problem of deriving the spectrum of the light scattered by a driven two-level atom is based on a master equation treatment on the atom-field interaction. Labeling $\ket{a}$ the ground state and $\ket{b}$ the excited state, one constructs the atomic density matrix
\begin{equation} \label{eq:AtomDensity}
\hat{\rho}_A\left(t\right)=\alpha\left(t\right)\ket{a}\bra{a}+\beta\left(t\right)\ket{b}\bra{b}+\gamma\left(t\right)\ket{a}\bra{b}+\gamma^*\left(t\right)\ket{b}\bra{a}
\end{equation}
from which a vector is constructed
\begin{equation} \label{eq:AtomDensityVector}
\bm{\rho}_A\left(t\right)=\left[\beta\left(t\right),\gamma\left(t\right),\gamma^*\left(t\right)\right]
\end{equation}
out of three of the four coefficients ($\alpha\left(t\right)$ is determined through $\alpha\left(t\right)=1-\beta\left(t\right)$). One then writes \cite{Mollow} the master equation
\begin{widetext}
\begin{equation} \label{eq:MasterEq}
\frac{\mathrm{d}\bm{\rho}_A}{\mathrm{d}t}=
\begin{bmatrix}
-\Gamma&\frac{\mathrm{i}}{2}\Omega^*&-\frac{\mathrm{i}}{2}\Omega\\
\mathrm{i}\Omega&-\mathrm{i}\Delta-\tfrac{1}{2}\Gamma&0\\
-\mathrm{i}\Omega^*&0&\mathrm{i}\Delta-\tfrac{1}{2}\Gamma-
\end{bmatrix}
\bm{\rho}_A\left(t\right)+
\begin{bmatrix}
0\\
-\frac{\mathrm{i}}{2}\Omega\\
\frac{\mathrm{i}}{2}\Omega^*
\end{bmatrix},
\end{equation}
\end{widetext}
where we recall that $\Delta=\omega_L-\omega_0$ is the laser-atom detuning, $\Gamma$ is the natural linewidth of the excited atomic level $\ket{b}$, and $\Omega$ is the Rabi frequency. Mollow's result was based on finding the steady-state solution to this equation, which is given by
\begin{subequations} \label{eq:SteadyState}
\begin{align}
\alpha_{\mathrm{SS}}&=\frac{4\Delta^2+\Gamma^2+\left|\Omega\right|^2}{4\Delta^2+\Gamma^2+2\left|\Omega\right|^2},\\
\beta_{\mathrm{SS}}&=\frac{\left|\Omega\right|^2}{4\Delta^2+\Gamma^2+2\left|\Omega\right|^2},\\
\gamma_{\mathrm{SS}}&=\frac{\left|\Omega\right|\left(2\Delta-\mathrm{i}\Gamma\right)}{4\Delta^2+\Gamma^2+2\left|\Omega\right|^2}
\end{align}
\end{subequations}
and $\gamma_{\mathrm{SS}}^*$ trivially obtained from (\ref{eq:SteadyState}). In this steady state, the coefficients of the atomic density matrix are constant but higher-order correlation functions are still dynamical quantities (sometimes referred to as quantum fluctuations around that steady state). It is the case of the correlation functions from which the scattered spectrum can be computed. Once the steady state has been established, one uses the quantum regression theorem \cite{LaxRegTh} to deduce the two-time expectation values $\left\langle\hat{\sigma}_+\left(0\right)\hat{\sigma}_-\left(t\right)\right\rangle$, the Fourier-Laplace transform of which yields the scattered spectrum. By making use of the quantum regression theorem, and after computing the Fourier-Laplace transform, Mollow found \cite{Mollow,JeKeLong}
\begin{multline} \label{eq:MollowFull}
S\left(\omega\right)=2\pi\left|\gamma_{\mathrm{SS}}\right|^2\delta\left(\omega-\omega_L\right)+16\,\beta_{\mathrm{SS}}\,\Gamma\,\left|\Omega\right|^2\\
\times\left[\frac{\left(\omega-\omega_L\right)^2+\left(\frac{\left|\Omega\right|^2}{2}+\Gamma^2\right)}{a_0\left(\omega\right)+a_2\left(\omega\right)\Gamma^2+a_4\left(\omega\right)\Gamma^4+a_6\left(\omega\right)\Gamma^6}\right]
\end{multline}
with the functions $a_i$ given by
\begin{subequations} \label{eq:Xa}
\begin{align}
a_0\left(\omega\right)&=16\left[\Delta^2+\left|\Omega\right|^2-\left(\omega-\omega_L\right)^2\right]^2\left(\omega-\omega_L\right)^2,\\
a_2\left(\omega\right)&=4\left[6\left(\omega-\omega_L\right)^2-2\left(3\Delta^2-\left|\Omega\right|^2\right)\left(\omega-\omega_L\right)^2\right.\nonumber\\
&\left.+\left(2\Delta^2+\left|\Omega\right|^2\right)^2\right],\\
a_4\left(\omega\right)&=8\Delta^2+4\left|\Omega\right|^2+9\left(\omega-\omega_L\right)^2,\\
a_6\left(\omega\right)&=1.
\end{align}
\end{subequations}
We can see in the light of (\ref{eq:MollowFull}) why the term proportional to the Dirac delta centered at the laser frequency is called the coherent spectrum: it is proportional to the square modulus of the coherence of the atomic density matrix (in the steady-state regime). The incoherent spectrum, on the other hand, is proportional to the population of the excited level: in the strong-field (high-intensity) regime, where the incoherent scattered spectrum splits into the three Mollow peaks, the atom has a non-negligible probability to be found in its excited state. A thorough analysis of the limiting cases of the Mollow spectrum (\ref{eq:MollowFull}) can be found in \cite{Mollow,JeKeLong}, and we need not repeat it here. Rather, we turn to our main point: the QED derivation of the master equation.

\section{Master equation in the resolvent formalism} \label{sec:ResoForma}

\subsection{General formalism} \label{subsec:Formalism}

We will now show how to confirm the results yielded by a quantum-optical treatment of the problem; by using the framework of the resolvent operator. For a system described by Hamiltonian $\hat{H}$, the resolvent operator is given by $\hat{G}\left(z\right)=\left(z-\hat{H}\right)^{-1}$ and is thus a function of a complex argument. It is well known, and easily understood, that $\hat{G}$ has singularities when $z$ is equal to an eigenvalue of the Hamiltonian $\hat{H}$. The Hamiltonian $\hat{H}$ might be chosen, in the case of QED, to be non-relativistic (as in our case) or relativistic. It is split into its free part $\hat{H}_0$ and its interacting part $\hat{V}$. The analytical structure of $\hat{G}$ in terms of the spectrum of the Hamiltonian provides one with all the information that is needed, in principle, to solve exactly for the dynamics of a given quantum system \cite{CohenQED2,FacchiPhD}.

We will work in the two-dimensional Hilbert subspace consisting of $\ket{a;\left(\gamma_L\right)^N}$ and $\ket{b;\left(\gamma_L\right)^{N-1}}$. We remind the reader that $\ket{a}$ is the atomic ground state and $\ket{b}$ the excited state; while $\gamma_L$ refers to a laser photon, constructed as a usual \cite{Ryder} wave packet
\begin{equation} \label{eq:LaserPhoton}
\ket{\gamma_L}=\sum_{\lambda=\pm}\int\tilde{\mathrm{d}k}\,f_{\left(\lambda\right)}\left(\mathbf{k}\right)\hat{a}_{\left(\lambda\right)}^\dagger\left(\mathbf{k}\right)\ket{0}
\end{equation}
where it is understood that $f_{\left(\lambda\right)}$ is heavily peaked around the laser frequency $\omega_L/c$. The vacuum state of the electromagnetic field is $\ket{0}$, and $\hat{a}_{\left(\lambda\right)}^\dagger\left(\mathbf{k}\right)$ is the creation operator for a photon of helicity $\lambda$ and wave vector $\mathbf{k}$. The invariant differential volume element on the light cone is
\begin{equation} \label{eq:InDaTilde}
\tilde{\mathrm{d}k}\equiv\frac{\mathrm{d}\mathbf{k}}{2\left(2\pi\right)^3\left|\mathbf{k}\right|}.
\end{equation}
Let us define $E_{aN}=E_a+N\hbar\omega_L$ and $E_{b\left(N-1\right)}=E_b+\left(N-1\right)\hbar\omega_L$. The atomic resonance frequency is $\hbar\omega_0=E_b-E_a$. Computing the matrix elements of the resolvent operator in this two-dimensional subspace will allow us to reconstruct the master equation ({\ref{eq:MasterEq}) rigorously. In the subspace of interest, these matrix elements are given by \cite{CohenQED2}:
\begin{multline} \label{eq:GElem}
\left[
\begin{array}{cc}
G_{aN}\left(z\right)&G_{aNb\left(N-1\right)}\left(z\right)\\
G_{b\left(N-1\right)aN}\left(z\right)&G_{b\left(N-1\right)}\left(z\right)
\end{array}
\right]=\frac{1}{\mathscr{D}_N\left(z\right)}\\
\times\left[
\begin{array}{cc}
z-E_{b\left(N-1\right)}-R_{b\left(N-1\right)}\left(z\right)&R_{aNb\left(N-1\right)}\left(z\right)\\
R_{b\left(N-1\right)aN}\left(z\right)&z-E_{aN}-R_{aN}\left(z\right)
\end{array}
\right],
\end{multline}
where the denominator is given by the determinant
\begin{multline} \label{eq:Determinant}
\mathscr{D}_N\left(z\right)\equiv\left(z-E_{b\left(N-1\right)}-R_{b\left(N-1\right)}\left(z\right)\right)\\
\times\left(z-E_{aN}-R_{aN}\left(z\right)\right)\\
-R_{aNb\left(N-1\right)}\left(z\right)R_{b\left(N-1\right)aN}\left(z\right),
\end{multline}
and $R$ refers to the level-shift operator, which is given by the Rayleigh-Schr\"{o}dinger expansion
\begin{equation} \label{eq:LevelShift}
\hat{R}\left(z\right)=\hat{V}+\hat{V}\frac{\hat{Q}}{z-\hat{H}_0}\hat{V}+\hat{V}\frac{\hat{Q}}{z-\hat{H}_0}\hat{V}\frac{\hat{Q}}{z-\hat{H}_0}\hat{V}+\ldots
\end{equation}
Here, $\hat{Q}$ is the projector over all possible quantum states of the system, except the two states of our subspace of reference:
\begin{equation} \label{eq:QProj}
\hat{Q}=\hat{\idmx}-\ket{a;\left(\gamma_L\right)^N}\bra{a;\left(\gamma_L\right)^N}-\ket{b;\left(\gamma_L\right)^{N-1}}\bra{b;\left(\gamma_L\right)^{N-1}}.
\end{equation}
The operator $\hat{V}$, finally, is the interaction Hamiltonian of the system. The quantities
\begin{subequations} \label{eq:RElements}
  \begin{align}
    R_{aN}\left(z\right)&\equiv\bra{a;\left(\gamma_L\right)^N}\hat{R}\left(z\right)\ket{a;\left(\gamma_L\right)^N},\\
    R_{b\left(N-1\right)}\left(z\right)&\equiv\bra{b;\left(\gamma_L\right)^{N-1}}\hat{R}\left(z\right)\ket{b;\left(\gamma_L\right)^{N-1}}
  \end{align}
\end{subequations}
give the radiative shifts of the levels $a$ and $b$, both due to their interaction with the photon vacuum and with the laser photons. The off-diagonal matrix elements of the level-shift operator also include, in principle, radiative shifts, but, in what follows, we will make the usual \cite{CohenQED2} approximation
\begin{subequations} \label{eq:OffDiagShift}
\begin{align}
R_{aNb\left(N-1\right)}\left(z\right)&\equiv\bra{a;\left(\gamma_L\right)^N}\hat{R}\left(z\right)\ket{b;\left(\gamma_L\right)^{N-1}}\nonumber\\
&\simeq\bra{a;\left(\gamma_L\right)^N}\hat{V}\ket{b;\left(\gamma_L\right)^{N-1}},\\
R_{b\left(N-1\right)aN}\left(z\right)&\equiv\bra{b;\left(\gamma_L\right)^{N-1}}\hat{R}\left(z\right)\ket{a;\left(\gamma_L\right)^N}\nonumber\\
&\simeq\bra{b;\left(\gamma_L\right)^{N-1}}\hat{V}\ket{a;\left(\gamma_L\right)^N}.
\end{align}
\end{subequations}
This approximation is, as a matter of fact, a strict consequence of the rotating wave approximation: if we consider that there can be no transitions whereby the atom absorbs a laser photon while going from the excited state $\ket{b}$ to the ground state $\ket{a}$, and, reciprocally, no transitions whereby the atom emits a photon while going from the ground to the excited state, then (\ref{eq:OffDiagShift}) is strictly correct. As such, in the rotating wave approximation, the off-diagonal elements of the level shift operator (\ref{eq:LevelShift}) are fully encapsulated by the lowest-order approximation in the interaction Hamiltonian. We write
\begin{equation} \label{eq:OffDLS}
R_{aN\,b\left(N-1\right)}=R_{b\left(N-1\right)\,aN}^*=\sqrt{N}\bra{a;\gamma_L}\hat{V}\ket{b;0}\equiv\sqrt{N}V
\end{equation}
where we introduced $V\equiv\bra{a;\gamma_L}\hat{V}\ket{b;0}$. Here we recognize an expression that is the (complex) Rabi frequency $\Omega\equiv2\sqrt{N}V/\hbar$.

In the small level-shift approximation, where the matrix elements of $\hat{R}$ are much smaller than the eigenstates of the free Hamiltonian $\hat{H}_0$, the poles of the resolvent operator (that is, the zeroes of the determinant (\ref{eq:Determinant})) are given \cite{CohenQED2} by
\begin{multline} \label{eq:PolePM}
z_\pm=\frac{1}{2}\left[E_a+E_b+\left(2N-1\right)\hbar\omega_L\right.\vphantom{R_{cM}\left(z_0\right)}\\
\left.+R_{aN}\left(z_0\right)+R_{b\left(N-1\right)}\left(z_0\right)\right]\\
\pm\frac{1}{2}\sqrt{\left[\Delta+R_{aN}\left(z_0\right)-R_{b\left(N-1\right)}\left(z_0\right)\right]^2+\left|\Omega\right|^2}.
\end{multline}
Here the reference energy $z_0$ is the average energy of the states $\ket{a;\left(\gamma_L\right)^N}$ and $\ket{b;\left(\gamma_L\right)^{N-1}}$:
\begin{equation} \label{eq:HalfEn}
z_0=\frac{1}{2}\left(E_a+E_b+\left(2N-1\right)\hbar\omega_L\right).
\end{equation}
Beyond the small level-shift approximation, the equation that determines the poles is a self-consistent one, and cannot be solved \emph{a priori}. In the present small level-shift case, we can go on to determine the level shifts of $E_a$ and $E_b$ for the values (\ref{eq:PolePM}) of the complex argument of the resolvent operator.

\subsection{Derivation of the master equation from the resolvent matrix} \label{eq:ResoReso}

Before we go on to computing the diagonal elements of the level-shift operator, we show how to derive a master equation from the current resolvent formalism. We make use of the relation between the time-evolution operator and the Green's operator (resolvent) for the atom-field system, namely \cite{CohenQED2}
\begin{equation} \label{eq:EtoU}
\hat{U}\left(t\right)=\frac{1}{2\pi\mathrm{i}}\int_{\mathcal{C}_++\mathcal{C}_-}\mathrm{d}z\,\mathrm{e}^{-\frac{\mathrm{i}}{\hbar}zt}\,\hat{G}\left(z\right).
\end{equation}
Here, the integration contour in the complex plane is given by the junction of two lines, $\mathcal{C}_+$ and $\mathcal{C}_-$, two horizontal lines situated, respectively, just above and just below the real axis, and followed from right to left for $\mathcal{C}_+$, and from left to right for $\mathcal{C}_-$. From the matrix elements (\ref{eq:GElem}), we can thence deduce the matrix elements of the evolution operator between the two states of the subspace of interest. From Equations~(\ref{eq:GElem}) to (\ref{eq:EtoU}), we derive (keeping in mind that the detuning is defined through $\Delta\equiv\omega_L-\omega_0$)
\begin{widetext}
\begin{subequations} \label{eq:UElem}
\begin{equation} \label{eq:DiagA}
\begin{aligned} [b]
U_a\left(t\right)\equiv\bra{a;\left(\gamma_L\right)^N}\hat{U}\left(t\right)\ket{a;\left(\gamma_L\right)^N}&=\mathrm{e}^{-\frac{\mathrm{i}}{\hbar}z_+t}\left(\frac{z_+-E_{b\left(N-1\right)}-R_{b\left(N-1\right)}\left(z_+\right)}{z_+-z_-}\right)\\
&+\mathrm{e}^{-\frac{\mathrm{i}}{\hbar}z_-t}\left(\frac{z_--E_{b\left(N-1\right)}-R_{b\left(N-1\right)}\left(z_-\right)}{z_--z_+}\right)\\
&=\frac{1}{2}\left[\left(1+\frac{\Delta+R_{aN}\left(z_0\right)+R_{b\left(N-1\right)}\left(z_0\right)-2R_{b\left(N-1\right)}\left(z_+\right)}{\sqrt{\left[\Delta+R_{aN}\left(z_0\right)+R_{b\left(N-1\right)}\left(z_0\right)\right]^2+\left|\Omega\right|^2}}\right)\mathrm{e}^{-\frac{\mathrm{i}}{\hbar}z_+t}\right.\\
&\left.\hspace{17.5pt}+\left(1-\frac{\Delta+R_{aN}\left(z_0\right)+R_{b\left(N-1\right)}\left(z_0\right)-2R_{b\left(N-1\right)}\left(z_-\right)}{\sqrt{\left[\Delta+R_{aN}\left(z_0\right)+R_{b\left(N-1\right)}\left(z_0\right)\right]^2+\left|\Omega\right|^2}}\right)\mathrm{e}^{-\frac{\mathrm{i}}{\hbar}z_-t}\right],\\
\end{aligned}
\end{equation}
\begin{equation} \label{eq:DiagB}
\begin{aligned} [b]
U_b\left(t\right)\equiv\bra{b;\left(\gamma_L\right)^{N-1}}\hat{U}\left(t\right)\ket{b;\left(\gamma_L\right)^{N-1}}&=\mathrm{e}^{-\frac{\mathrm{i}}{\hbar}z_+t}\left(\frac{z_+-E_{aN}-R_{aN}\left(z_+\right)}{z_+-z_-}\right)+\mathrm{e}^{-\frac{\mathrm{i}}{\hbar}z_-t}\left(\frac{z_--E_{aN}-R_{aN}\left(z_-\right)}{z_--z_+}\right)\\
&=\frac{1}{2}\left[\left(1-\frac{\Delta-R_{aN}\left(z_0\right)-R_{b\left(N-1\right)}\left(z_0\right)+2R_{aN}\left(z_+\right)}{\sqrt{\left[\Delta+R_{aN}\left(z_0\right)+R_{b\left(N-1\right)}\left(z_0\right)\right]^2+\left|\Omega\right|^2}}\right)\mathrm{e}^{-\frac{\mathrm{i}}{\hbar}z_+t}\right.\\
&\left.\hspace{17.5pt}+\left(1+\frac{\Delta-R_{aN}\left(z_0\right)-R_{b\left(N-1\right)}\left(z_0\right)+2R_{aN}\left(z_-\right)}{\sqrt{\left[\Delta+R_{aN}\left(z_0\right)+R_{b\left(N-1\right)}\left(z_0\right)\right]^2+\left|\Omega\right|^2}}\right)\mathrm{e}^{-\frac{\mathrm{i}}{\hbar}z_-t}\right],\\
\end{aligned}
\end{equation}
\begin{equation} \label{eq:OffDiag}
\begin{aligned} [b]
U_{ab}\left(t\right)\equiv\bra{a;\left(\gamma_L\right)^N}\hat{U}\left(t\right)\ket{b;\left(\gamma_L\right)^{N-1}}&=\mathrm{e}^{-\frac{\mathrm{i}}{\hbar}z_+t}\left(\frac{\frac{1}{2}\Omega}{z_+-z_-}\right)+\mathrm{e}^{-\frac{\mathrm{i}}{\hbar}z_-t}\left(\frac{\frac{1}{2}\Omega}{z_--z_+}\right)\\
&=\frac{1}{2}\Omega\left[\frac{\mathrm{e}^{-\frac{\mathrm{i}}{\hbar}z_+t}-\mathrm{e}^{-\frac{\mathrm{i}}{\hbar}z_-t}}{\sqrt{\left[\Delta+R_{aN}\left(z_0\right)+R_{b\left(N-1\right)}\left(z_0\right)\right]^2+\left|\Omega\right|^2}}\right].\\
\end{aligned}
\end{equation}\end{subequations}
\end{widetext}
We can then construct the master equation, this time for the atom-field density matrix
\begin{equation} \label{eq:AtomFieldDensity}
\begin{aligned} [b]
\hat{\rho}\left(t\right)&=A\left(t\right)\ket{a;\left(\gamma_L\right)^N}\bra{a;\left(\gamma_L\right)^N}\\
&+B\left(t\right)\ket{b;\left(\gamma_L\right)^{N-1}}\bra{b;\left(\gamma_L\right)^{N-1}}\\
&+C\left(t\right)\ket{a;\left(\gamma_L\right)^N}\bra{b;\left(\gamma_L\right)^{N-1}}\\
&+C^*\left(t\right)\ket{b;\left(\gamma_L\right)^{N-1}}\bra{a;\left(\gamma_L\right)^N}.
\end{aligned}
\end{equation}
To do this, once again we construct a vector
\begin{equation} \label{eq:AtomDFieldensityVector}
\bm{\rho}\left(t\right)=\left[B\left(t\right),C\left(t\right),C^*\left(t\right)\right]
\end{equation}
out of three of the four coefficients (with $A\left(t\right)=1-B\left(t\right)$), and for these coefficients we write the equation
\begin{subequations} \label{eq:UGalore}
\begin{equation} \label{eq:RhoLater}
\bm{\rho}\left(t+t_0\right)=\mathscr{U}\left(t\right)\bm{\rho}\left(t_0\right)+\mathbf{U}\left(t\right),
\end{equation}
with
\begin{widetext}
\begin{equation} \label{eq:UMatrix}
\mathscr{U}\left(t\right)=
\left[
\begin{array}{ccc}
-U_{ab}^*\left(t\right)U_{ab}+U_b^*\left(t\right)U_b\left(t\right)&U_{ab}^*\left(t\right)U_b\left(t\right)&U_b^*\left(t\right)U_{ab}\left(t\right)\\
-U_a^*\left(t\right)U_{ab}\left(t\right)+U_{ba}^*\left(t\right)U_b\left(t\right)&U_a^*\left(t\right)U_b\left(t\right)&U_{ba}^*\left(t\right)U_{ab}\left(t\right)\\
-U_{ab}^*\left(t\right)U_a\left(t\right)+U_b^*\left(t\right)U_{ba}\left(t\right)&U_{ab}^*\left(t\right)U_{ba}\left(t\right)&U_b^*\left(t\right)U_a\left(t\right)
\end{array}
\right]
\end{equation}
\end{widetext}
and
\begin{equation} \label{qe:UVector}
\mathbf{U}\left(t\right)=
\left[
\begin{array}{c}
U_{ab}^*\left(t\right)U_{ab}\left(t\right)\\
U_a^*\left(t\right)U_{ab}\left(t\right)\\
U_{ab}^*\left(t\right)U_a\left(t\right)
\end{array}
\right].
\end{equation}
\end{subequations}
We can then Taylor-expand the evolution equation at the first order in $t$, yielding
\begin{equation} \label{eq:Derivative}
\frac{\mathrm{d}\bm{\rho}}{\mathrm{d}t}=\lim_{\epsilon\to0}\frac{1}{\epsilon}\left[\left(\mathscr{U}\left(\epsilon\right)-\idmx\right)\bm{\rho}\left(t\right)+\mathbf{U}\left(\epsilon\right)\right].
\end{equation}
This is the abstract, general form of the master equation. It remains to be verified that it matches the quantum optical equation (\ref{eq:MasterEq}) of Mollow. Substituting (\ref{eq:UElem}) into (\ref{eq:UGalore}) yields cumbersome expressions, which we will not reproduce here as they are of limited relevance. It is better, at this stage, to turn to an explicit determination of the level shifts.

\subsection{Determination of the level shifts} \label{subsec:LevelShifts}

It then becomes important to obtain the radiative energy shifts of the quantum states $\ket{a;\left(\gamma_L\right)^N}$ and $\ket{b;\left(\gamma_L\right)^{N-1}}$ under the effect of the interaction Hamiltonian. It is well known \cite{FoxMulder,JeKeLong} that, under the influence of a strong driving laser field, a two-level system exhibits the so-called dressed states, with energy $E_\pm=\left(E_a+E_b+\left(2N-1\right)\hbar\omega_L\right)/2\pm\left(1/2\right)\sqrt{\left|\Omega\right|^2+\Delta^2}$. As can be seen from (\ref{eq:PolePM}), this radiative shift, which we could call the Rabi dressing shift, has already been taken into account by our working in the two-dimensional subspace consisting of $\ket{a;\left(\gamma_L\right)^N}$ and $\ket{b;\left(\gamma_L\right)^{N-1}}$, without any need to include the diagonal matrix elements of the level-shift operator. The Rabi dressing shift corresponds to a process whereby the atom repeatedly undergoes transitions between its ground state $a$ and excited state $b$ while absorbing and emitting laser photons $\gamma_L$. In our formalism, this is taken into account at the outset. Let us now turn to the explicit determination of the shifts, as prescribed by (\ref{eq:RElements}). We have so far performed the rotating wave approximation, and for consistency will continue doing so here. We need to compute the diagonal matrix elements of the level-shift operator (\ref{eq:LevelShift}). Starting with the easier case of the expectation value in state $\ket{a;\left(\gamma_L\right)^N}$, we note that, at the rotating wave approximation, only photon absorption can excite the atom. However, from the reference state $\ket{a;\left(\gamma_L\right)^N}$, only laser photons are available for absorption, and this absorption process takes the system to $\ket{b;\left(\gamma_L\right)^{N-1}}$, which cannot contribute because of the projector $\hat{Q}$ defined by (\ref{eq:QProj}). Hence, the level shift is zero \footnote{We note in passing that, had we chosen a one-dimensional Hilbert subspace of reference in which to apply the resolvent formalism, consisting solely of the sole state $\ket{a;\left(\gamma_L\right)^N}$, the projector $\hat{Q}'=\hat{\idmx}-\ket{a;\left(\gamma_L\right)^N}\bra{a;\left(\gamma_L\right)^N}$ would not eliminate the contribution of $\ket{b;\left(\gamma_L\right)^{N-1}}$, and the Rabi dressing shift would hence be obtained by computing the diagonal matrix element of the level-shift operator (instead of being obtained, as in our treatment, by a privileged treatment of the transition $\ket{a;\left(\gamma_L\right)^N}\leftrightarrow\ket{b;\left(\gamma_L\right)^{N-1}}$). Of course, such a treatment fails to describe the effect of the laser-atom interaction on the state $\ket{b;\left(\gamma_L\right)^{N-1}}$ in sufficient detail.}. Now turn to the expectation value in state $\ket{b;\left(\gamma_L\right)^{N-1}}$. Identically, the projector $\hat{Q}$ roots out the contribution of Rabi flopping to the shift, consistent with the fact that this contribution has intrinsically been included in the treatment. However, this time, a photon of arbitrary characteristics can be emitted and reabsorbed from and to state $\ket{b;\left(\gamma_L\right)^{N-1}}$. Interestingly, it needs to be reabsorbed at the last step of the process, as $\ket{b;\left(\gamma_L\right)^{N-1}}$ cannot be an intermediate state. Explicitly, only a single series of diagrams contributes, the first three of which are represented on Fig.~\ref{fig:Shift}. The series is given by the sum over all (integer) numbers of Rabi oscillations inside the self-energy loop of level $b$:
{\allowdisplaybreaks
\begin{widetext}
\begin{align} \label{eq:GeneralShift}
&\bra{b;\left(\gamma_L\right)^{N-1}}\hat{R}\left(z\right)\ket{b;\left(\gamma_L\right)^{N-1}}\nonumber\\
&=\bra{b;\left(\gamma_L\right)^{N-1}}\hat{V}\frac{Q}{z-\hat{H}_0}\hat{V}\ket{b;\left(\gamma_L\right)^{N-1}}+\bra{b;\left(\gamma_L\right)^{N-1}}\hat{V}\frac{Q}{z-\hat{H}_0}\hat{V}\frac{Q}{z-\hat{H}_0}\hat{V}\frac{Q}{z-\hat{H}_0}\hat{V}\ket{b;\left(\gamma_L\right)^{N-1}}+\ldots\nonumber\\
&=\sum_{\lambda=\pm}\int\tilde{\mathrm{d}q}\left|G_{ab\left(\lambda\right)}\left(\mathbf{q}\right)\right|^2\frac{1}{z-E_a-\left(N-1\right)\hbar\omega_L-\hbar c\left|\mathbf{q}\right|}\nonumber\\
&\hspace{112.5pt}\times\left[1+\left(N-1\right)\left|V\right|^2\frac{1}{z-E_a-\left(N-1\right)\hbar\omega_L-\hbar c\left|\mathbf{q}\right|}\frac{1}{z-E_b-\left(N-2\right)\hbar\omega_L-\hbar c\left|\mathbf{q}\right|}+\ldots\right]\nonumber\\
&=\sum_{\lambda=\pm}\int\tilde{\mathrm{d}q}\left|G_{ab\left(\lambda\right)}\left(\mathbf{q}\right)\right|^2\frac{1}{z-E_a-\left(N-1\right)\hbar\omega_L-\hbar c\left|\mathbf{q}\right|}\nonumber\\
&\hspace{112.5pt}\times\frac{1}{1-\left(N-1\right)\left|V\right|^2\left(z-E_a-\left(N-1\right)\hbar\omega_L-\hbar c\left|\mathbf{q}\right|\right)^{-1}\left(z-E_b-\left(N-2\right)\hbar\omega_L-\hbar c\left|\mathbf{q}\right|\right)^{-1}}\nonumber\\
&=\sum_{\lambda=\pm}\int\tilde{\mathrm{d}q}\left|G_{ab\left(\lambda\right)}\left(\mathbf{q}\right)\right|^2\frac{\left(z-E_b-\left(N-2\right)\hbar\omega_L-\hbar c\left|\mathbf{q}\right|\right)}{\left(z-E_a-\left(N-1\right)\hbar\omega_L-\hbar c\left|\mathbf{q}\right|\right)\left(z-E_b-\left(N-2\right)\hbar\omega_L-\hbar c\left|\mathbf{q}\right|\right)-\left(N-1\right)\left|V\right|^2}.
\end{align}
\end{widetext}
}
\begin{figure}[t]
\includegraphics[width=0.9 \textwidth]{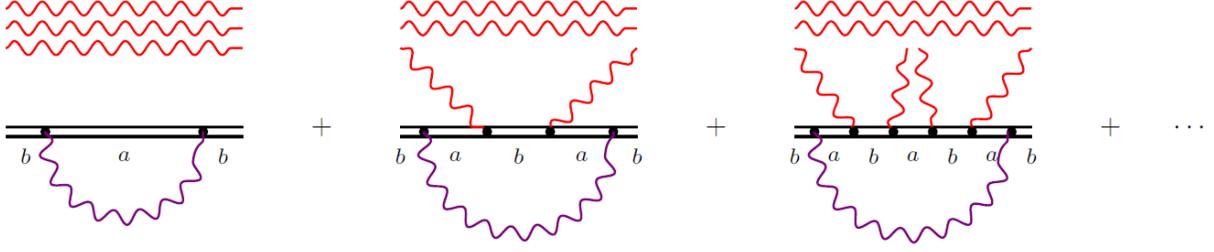}
\caption{Diagrams corresponding to the processes contributing to the energy shift of the state $\ket{b;\left(\gamma_L\right)^{N-1}}$. The series continues by inserting further Rabi oscillations (absorption and emission of laser photons) between the ground state $a$ and he excited state $b$, inside the self-energy loop. The double line represents the bound electron, red wavy lines are laser photons, while the violet wavy line is the intermediate (self-energy loop) photon.
\label{fig:Shift}}
\end{figure}
Here, in order to be able to carry out the resummation, we have ignored, as is customary \cite{CohenQED2}, nested loop, and overlapping loop, diagrams. Now, as we learned in Sec.~\ref{subsec:Formalism}, we will need to take the value of this shift for $z=z_0$ given by (\ref{eq:HalfEn}), as well as for $z=z_\pm$ given by (\ref{eq:PolePM}). Notice, first, that, by definition, the natural shift of the excited level $b$ is given by
\begin{equation} \label{eq:NaturalShift}
N_b=\sum_{\lambda=\pm}\int\tilde{\mathrm{d}q}\,\frac{\left|G_{ab\left(\lambda\right)}\left(\mathbf{q}\right)\right|^2}{\hbar\left(\omega_0-c\left|\mathbf{q}\right|\right)+\mathrm{i}\epsilon},
\end{equation}
where it is understood that the limit $\epsilon\rightarrow0$ is to be taken, allowing one to compute the real part (Lamb-type shift) and imaginary part (natural linewidth) of the shift with the help of the Sochocki-Plemelj theorem \cite{CohenQED2}. Here the coupling function is given by
\begin{equation} \label{eq:CouplingG}
G_{ab\left(\lambda\right)}\left(\mathbf{q}\right)=\bra{a;\left(\mathbf{q},\lambda\right)}\hat{V}\ket{b;0}.
\end{equation}
We deduce the natural linewidth $\Gamma=-2\,\mathfrak{Im}\,N_b$:
\begin{equation} \label{eq:NaturalWidth}
\Gamma=2\pi\sum_{\lambda=\pm}\int\tilde{\mathrm{d}q}\,\frac{\left|G_{ab\left(\lambda\right)}\left(\mathbf{q}\right)\right|^2}{\hbar c}\delta\left(\left|\mathbf{q}\right|-\frac{\omega_0}{c}\right).
\end{equation}
Now, we investigate how interaction with the laser field modifies this natural linewidth. For that we need to compute the shift (\ref{eq:GeneralShift}), as explained just above, for $z=z_0$ and $z=z_\pm$. We will focus in what follows on the imaginary part of the shift, considering that the real part can simply be reabsorbed, as is common, in the value $E_b$ of the energy of level $b$. We have, from (\ref{eq:HalfEn}) and (\ref{eq:GeneralShift}), the shift
\begin{widetext}
\begin{equation*}
\bra{b;\left(\gamma_L\right)^{N-1}}\hat{R}\left(z_0\right)\ket{b;\left(\gamma_L\right)^{N-1}}=\sum_{\lambda=\pm}\int\tilde{\mathrm{d}q}\left|G_{ab\left(\lambda\right)}\left(\mathbf{q}\right)\right|^2\frac{\frac{1}{2}\left[\Delta+2\hbar\left(\omega_L-c\left|\mathbf{q}\right|\right)\right]}{\frac{1}{2}\left[-\Delta+2\hbar\left(\omega_L-c\left|\mathbf{q}\right|\right)\right]\frac{1}{2}\left[\Delta+2\hbar\left(\omega_L-c\left|\mathbf{q}\right|\right)\right]-\left(N-1\right)\left|V\right|^2}
\end{equation*}
and the zeroes of the denominator can be found for
\begin{equation}
\left|\mathbf{q}\right|_\pm=\frac{\omega_L}{c}+\frac{1}{2\hbar c}\sqrt{\Delta^2+4\left(N-1\right)\left|V\right|^2},
\end{equation}
from which we deduce, after a few steps of algebra (and with the help of the Sochocki-Plemelj theorem), that the decay width for $z=z_0$ is
\begin{equation} \label{eq:DecayZero}
\Gamma_0=2\pi\sum_{\lambda=\pm}\int\tilde{\mathrm{d}q}\,\frac{\left|G_{ab\left(\lambda\right)}\left(\mathbf{q}\right)\right|^2}{2\hbar c}\sum_\pm\delta\left(\left|\mathbf{q}\right|-\frac{\omega_L}{c}\mp\frac{1}{2\hbar c}\sqrt{\Delta^2+4\left(N-1\right)\left|V\right|^2}\right)\left(1\mp\frac{\Delta}{\sqrt{\Delta^2+4\left(N-1\right)\left|V\right|^2}}\right).
\end{equation}
With a similar method, we can establish, on the basis of (\ref{eq:PolePM}) and (\ref{eq:GeneralShift}), the decay width for $z=z_\epsilon$ (with $\epsilon=+$ or $-$) as
\begin{multline} \label{eq:DecayPM}
\Gamma_\epsilon\simeq2\pi\sum_{\lambda=\pm}\int\tilde{\mathrm{d}q}\,\frac{\left|G_{ab\left(\lambda\right)}\left(\mathbf{q}\right)\right|^2}{2\hbar c}\sum_\pm\delta\left(\left|\mathbf{q}\right|-\frac{\omega_L}{c}\mp\frac{1}{2\hbar c}\sqrt{\Delta^2+4\left(N-1\right)\left|V\right|^2}-\frac{\epsilon}{2\hbar c}\sqrt{\Delta^2+4N\left|V\right|^2}\right)\\
\times\left(1\mp\frac{\Delta}{\sqrt{\Delta^2+4\left(N-1\right)\left|V\right|^2}}\right).
\end{multline}
\end{widetext}
Now, we specify that the atom-field interaction Hamiltonian is given by
\begin{equation} \label{eq:InterHam}
\hat{V}=\frac{e}{m_e}\hat{\mathbf{A}}\left(\hat{\mathbf{x}},t=0\right)\cdot\hat{\mathbf{p}}.
\end{equation}
The matrix elements of this Hamiltonian have been thoroughly studied, \emph{e.g.} by Seke \cite{Seke}. In our notation, for all (relevant \footnote{To the exclusion of electric dipole transitions between states that share the same principal quantum number. Such transitions, in any case, are very slow, and hence not compatible with the two-level approximation: the excited state will decay much more quickly to a third level. This remains true in the fully relativistic case \cite{GrantCoupling}.}) transitions of the electric dipole type, the coupling function as we defined it [see (\ref{eq:CouplingG})] varies very slowly except when the frequency becomes comparable to a cutoff frequency of the order of $\alpha^{-1}$ times the ($Z$-scaled) Hartree energy $E_C=\left(Z\alpha\right)^2 m_e c^2$. No frequency featured on the Dirac $\delta$ distributions of (\ref{eq:DecayZero}) and (\ref{eq:DecayPM}) approaches that order of magnitude. Indeed, we consider that the laser frequency $\omega_L$ is broadly of the same magnitude as the transition frequency $\omega_0$. The detuning, hence, is at most, also of the order $\omega_0$, but typically much smaller (as one often tries to achieve resonance $\omega_L=\omega_0$). The coupling strength(s) $\sqrt{N}\left|V\right|$ (and $\sqrt{N-1}\left|V\right|$), in turn, are much smaller than the atomic resonance frequency (both for experimental reasons, and because the two-level formalism with the attending rotating wave approximation would break down if it were not the case \cite{FloquetRWA}). As such, the couplings in (\ref{eq:DecayZero}) and (\ref{eq:DecayPM}) may be approximated by their value at the atomic transition frequency. Explicitly, we have
\begin{equation} \label{eq:FinalDecayZero}
\Gamma_0\simeq\Gamma
\end{equation}
where the calculation is different for, on the one hand, the case where $\Delta\ll\left|\Omega\right|$ or $\Delta\sim\left|\Omega\right|$ and, on the other hand, the case where $\Delta\gg\left|\Omega\right|$, but the result (\ref{eq:FinalDecayZero}) is the same. For $\Delta\ll\left|\Omega\right|$ or $\Delta\sim\left|\Omega\right|$, we also found that
\begin{subequations} \label{eq:FinalDecayPM}
  \begin{equation} \label{eq:LowDet}
    \Gamma_\pm\simeq\Gamma.
  \end{equation}
However, for large detuning $\Delta\gg\left|\Omega\right|$ compared to the Rabi frequency, we obtain the somewhat more involved results
  \begin{equation} \label{eq:HighDet}
    \Gamma_\pm\simeq\left[1+\theta\left(\pm\Delta\right)\left(\frac{\omega_L}{\omega_0}-1\right)\right]\Gamma,
  \end{equation}
\end{subequations}
where we see that either one of the $\Gamma_\pm=-2\mathfrak{Im}\bra{b;\left(\gamma_L\right)^{N-1}}\hat{R}\left(z_\pm\right)\ket{b;\left(\gamma_L\right)^{N-1}}$ may be modified in function of the sign of the detuning $\Delta$. We will therefore focus on reasonably small detunings such that $\Delta\ll\left|\Omega\right|$ or $\Delta\sim\left|\Omega\right|$ in the final steps of our derivation. For such detunings, the interaction with the driving field keeps the linewidth of level $b$ intact, to a very good approximation, even in the strong-field case.

\subsection{QED master equation and discussion} \label{eq:Final}

Let us restart from the matrix elements (\ref{eq:UElem}) of the evolution operator. Therein, we plug the results from the previous Sec.~\ref{subsec:LevelShifts}, namely
\begin{subequations} \label{eq:AllShifts}
\begin{align}
R_{aN}\left(z_0\right)=R_{aN}\left(z_\pm\right)&=0,\\
R_{b\left(N-1\right)}\left(z_0\right))=R_{b\left(N-1\right)}\left(z_\pm\right)&=-\frac{\mathrm{i}}{2}\Gamma,
\end{align}
\end{subequations}
which yields, from (\ref{eq:UElem}) and (\ref{eq:UGalore}), the matrix
\begin{subequations} \label{eq:WhatU}
\begin{equation} \label{eq:WhatUMatrix}
\lim_{\epsilon\to0}\frac{\mathscr{U}\left(\epsilon\right)-\idmx}{\epsilon}=
\begin{bmatrix}
-\Gamma&\frac{\mathrm{i}}{2}\Omega^*&-\frac{\mathrm{i}}{2}\Omega\\
\mathrm{i}\Omega&-\mathrm{i}\Delta-\tfrac{1}{2}\Gamma&0\\
-\mathrm{i}\Omega^*&0&\mathrm{i}\Delta-\tfrac{1}{2}\Gamma
\end{bmatrix},
\end{equation}
where we made use of
\begin{equation*}
\begin{aligned} [b]
\mathfrak{Im}\,z_++\mathfrak{Im}\,z_-&=\mathfrak{Im}\left(R_{aN}\left(z_0\right)+R_{b\left(N-1\right)}\left(z_0\right)\right)\\
&=-\frac{\Gamma}{2},
\end{aligned}
\end{equation*}
and the vector
\begin{equation} \label{qe:UVector}
\lim_{\epsilon\to0}\frac{\mathbf{U}\left(\epsilon\right)}{\epsilon}=
\begin{bmatrix}
0\\
-\frac{\mathrm{i}}{2}\Omega\\
\frac{\mathrm{i}}{2}\Omega^*
\end{bmatrix},
\end{equation}
\end{subequations}
which, on the basis of (\ref{eq:Derivative}), establishes that the atom-field density matrix obeys the Mollow master equation (\ref{eq:MasterEq}). As we have mentioned above, this is sufficient to establish the Mollow spectrum. We remember that, for large detunings $\Delta\gg\left|\Omega\right|$ compared to the Rabi frequency, the linewidth of the excited level $b$ is modified, which complicates the resulting master equation away from that of Mollow \cite{Mollow}. As such, we might expect corrections to the Mollow spectrum in that case, but this is outside the scope of the work presented here.

Our formalism allows for the inclusion of further radiative corrections to the Mollow spectrum. In our model, that of a two-level atom  at the rotating wave approximation, the only contribution to the shifts which we have neglected are those of nested self-energy loops for the excited level. We have opted, in a sense, for a treatment sufficiently involved to yield the Mollow spectrum, but no more. It would be very interesting indeed to extend our treatment to the many-level case, which allows more naturally for the inclusion of counter-rotating terms which have been excluded here. Indeed, in our treatment, the Rabi frequency is determined by the off-diagonal matrix element of the resolvent operator, and is strictly equal to the coupling strength between the atom and the laser. In the presence of further atomic levels, available for virtual transitions from either one of the two levels $a$ and $b$ featured in the addressed transition, higher-order corrections (for instance, of the polarization type) to the off-diagonal element are to be anticipated, yielding corrections to the Rabi frequency.

\section{Conclusion} \label{sec:Ccl}

The generalized description of resonant light-matter interactions which we have developed here allows for a natural inclusion of radiative corrections, already well understood in the case of weak-field excitations \cite{LowLineShape}, in the description of multiphoton processes. Specifically, we anticipate, as explained just above, that in a complete account of fluorescence spectra, not only the position of the emission lines and the radiative decay widths have to be corrected by QED radiative corrections, as would be the case in the weak-field limit \cite{LowLineShape}, but so does the Rabi frequency. More broadly, a formulation of resonant interactions between intense fields and atomic systems provides novel means of testing the validity of strong-field QED in a dynamical setting: indeed, while QED has been benchmarked to ultimate accuracy with respect to the static features of atoms and ions, such as transition energies or $g$-factors \cite{Sturm11}, the same may not be said when dynamical phenomena, \emph{e.g.} resonant photon scattering, are concerned. Laser spectroscopic experiments, for instance, would have the sufficient precision for observing dynamical QED phenomena, which are anticipated to play an enhanced role at higher atomic numbers, and, correspondingly, at higher atomic transition frequencies.

\section*{Acknowledgments}

We thank Stefano M.~Cavaletto and Emmanuel Lassalle for helpful discussions.

\end{document}